\documentclass[pdflatex,sn-mathphys-num]{sn-jnl}

\usepackage{graphicx}%
\usepackage{multirow}%
\usepackage{amsmath,amssymb,amsfonts}%
\usepackage{amsthm}%
\usepackage{mathrsfs}%
\usepackage[title]{appendix}%
\usepackage{xcolor}%
\usepackage{textcomp}%
\usepackage{manyfoot}%
\usepackage{booktabs}%
\usepackage{algorithm}
\usepackage{algpseudocode}
\usepackage{algorithm}%
\usepackage{algorithmicx}%
\usepackage{algpseudocode}%
\usepackage{listings}%
\usepackage{braket}

\theoremstyle{thmstyleone}%
\theoremstyle{thmstyletwo}%

\theoremstyle{thmstylethree}%

\begin{document}

\title[Article Title]{HOPSO: A Robust Classical Optimizer for VQE}


\author*[1]{\fnm{Ijaz Ahamed} \sur{Mohammad}}\email{fyziijaz@savba.sk}
\equalcont{These authors contributed equally to this work.}
\author[1]{\fnm{Yury} \sur{Chernyak}}\email{theofficialyury@gmail.com}
\author[1,2,3]{\fnm{Martin} \sur{Plesch}}\email{martin.plesch@savba.sk}
\equalcont{These authors contributed equally to this work.}

\affil*[1]{\orgname{Institute of Physics}, \orgaddress{\street{ Dúbravská cesta 5807/9}, \city{Bratislava}, \postcode{84511}, \country{Slovakia}}}
\affil[2]{\orgname{Matej Bel University}, \orgaddress{\street{Národná ulica 12}, \city{Banská Bystrica}, \postcode{97401},  \country{Slovakia}}}
\affil[3]{\orgname{Institute of Physics of Materials}, \orgaddress{\street{ Žižkova 22}, \city{Brno}, \postcode{616 62}, \country{Czech Republic}}}


\abstract{Variational Quantum Eigensolver (VQE) algorithm is one of few approaches where the hope  for near-term quantum advantage concentrates. 
However, they face challenges connected with measurement stochastic noise, barren plateaus, and optimization difficulties in periodic parameter spaces. 
While most of the efforts concentrates on optimizing the quantum part of the procedure, here we aim to enhance the classical optimization by utilizing a modified version of Harmonic Oscillator-based Particle Swarm Optimization (HOPSO).
By adapting its dynamics to respect the periodicity of quantum parameters and enhance noise resilience, we show its strengths on hydrogen (\textnormal{H}$_2$) and lithium hydride (LiH) molecules modeled as 4- and 8-qubit Hamiltonians.
HOPSO achieves competitive ground-state energy approximations and demonstrates improved robustness compared to COBYLA, Differential Evolution (DE), and standard Particle Swarm Optimization (PSO) methods in all situations and outperforms other methods under realistic noise conditions. These results suggest that a properly tailored classical part of VQE algorithms can tackle with current problems and gives hope for its scalability for larger systems. }

\keywords{Variational Quantum Eigensolver (VQE), Quantum Optimization, Harmonic Oscillator based PSO (HOPSO), Swarm Intelligence, Periodic Cost Functions, Quantum Noise Resilience, Quantum Chemistry Simulation, Noisy Intermediate-Scale Quantum (NISQ), Metaheuristic algorithms, Global Optimization, Barren Plateaus, Ansatz design, Jordan-Wigner Transformation}



\maketitle

\section{Introduction}\label{sec1}

Quantum computers hold the promise of solving certain problems more efficiently than their classical counterparts.
Among the most prominent frameworks for demonstrating this promise of near-term quantum advantage is the Variational Quantum Eigensolver (VQE), a hybrid quantum-classical algorithm used to approximate ground-state energies of molecular systems.
By variationally optimizing a parametrized quantum circuit, VQE estimates the minimum eigenvalue of a Hamiltonian using a quantum subroutine to evaluate expectation values and a classical optimizer to update circuit parameters~\cite{peruzzo2014variational, cerezo2021variational}.

Despite its potential, the scalability and robustness of VQE remain active areas of investigation.
Recent studies \cite{cerezo2021challenges,PhysRevResearch.2.043158,huggins2021efficient, mohammad2024meta} have highlighted key challenges that limit its practical performance: most notably, the emergence of \textit{barren plateaus}, which pose a fundamental obstacle to gradient-based optimization, the \textit{stochastic nature} of objective function evaluations due to quantum measurement noise, and the underlying \textit{periodic structure} of the parameter space.
Although the periodicity of quantum gates is not inherently problematic, many classical optimization strategies fail to account for this structure appropriately, often ignoring the fact that the parameters wrap around like angles (thereby leading them to search in straight lines instead of following the circular nature of rotational quantum gates).
This mismatch can lead to inefficient exploration, finding the local instead of global minima and failing to converge.

While VQE has recently faced criticism regarding its scalability and reliability~\cite{wang2023noise,cerezo2021challenges,mcclean2018barren}, we argue that part of this skepticism stems from underestimating the role of classical optimization.
In particular, many classical methods struggle not due to fundamental limits, but because they are not adapted to the unique structure of quantum variational landscapes and the intrinsic quantum feature of dynamic stochastic change of cost function due to measurement noise. 
Our work suggests that with appropriate design—especially physically motivated (and structurally-aware)  strategies like the HOPSO algorithm—classical optimization remains a powerful and underutilized tool in quantum algorithm development.

In this work, we build on our earlier development of the Harmonic Oscillator-based Particle Swarm Optimization (HOPSO) algorithm~\cite{10.1371/journal.pone.0326173}, originally proposed for solving classical global optimization problems.
We extend HOPSO to operate within the VQE framework, introducing specific modifications that allow it to navigate through periodic domains and noisy measurement evaluations effectively.
By adapting the attractor dynamics—i.e., the oscillatory motion of particles around the weighted center of personal and global best positions—and enforcing periodic boundary handling, we tailor HOPSO to the constraints of a quantum parameter space without sacrificing its dynamic search behavior.

We evaluate the modified algorithm on two standard molecular systems: hydrogen (H$_2$) and lithium hydride (LiH), encoded as 4- and 8-qubit Hamiltonians, respectively.
The performance of HOPSO is benchmarked against three widely used optimizers—COBYLA \cite{powell1994direct}, Differential Evolution (DE) \cite{storn1997differential}, and standard PSO \cite{kennedy1995particle}—under both noiseless and noisy conditions. Our results show that HOPSO offers consistent convergence, high precision, and robustness to quantum noise, particularly in more complex, high-dimensional scenarios such as that of lithium hydride.

This paper is organized as follows. Section~\ref{sec: background} provides background on VQE and outlines the core ideas of the original HOPSO method.
Section~\ref{sec: method} introduces the modifications made to adapt HOPSO to quantum optimization. Sections~\ref{sec4} and \ref{sec5} presents numerical experiments with H$_2$ and LiH, including analyses of performance under realistic noise.
We conclude in Section~\ref{sec:conclusion} with a discussion of findings, limitations, and future directions.

\section{Variational Quantum Eigensolver}\label{sec: background}

The Variational Quantum Eigensolver (VQE)~\cite{peruzzo2014variational} is a hybrid quantum-classical algorithm designed to estimate the ground-state energy of a given Hamiltonian.
It is particularly suited for near-term quantum devices due to its relatively low circuit depth and its reliance on classical optimization to guide quantum evaluations.
VQE operates by preparing a parametrized quantum state $\ket{\psi(\boldsymbol{\theta})}$ using a variational ansatz, typically composed of layers of parameterized rotation gates and entangling gates. The expectation value of the target Hamiltonian $H$ with respect to this state, $\langle \psi(\boldsymbol{\theta}) | H | \psi(\boldsymbol{\theta}) \rangle$, is then estimated through quantum measurements.

\begin{figure} 
    \centering
    \includegraphics[width=0.8\linewidth]{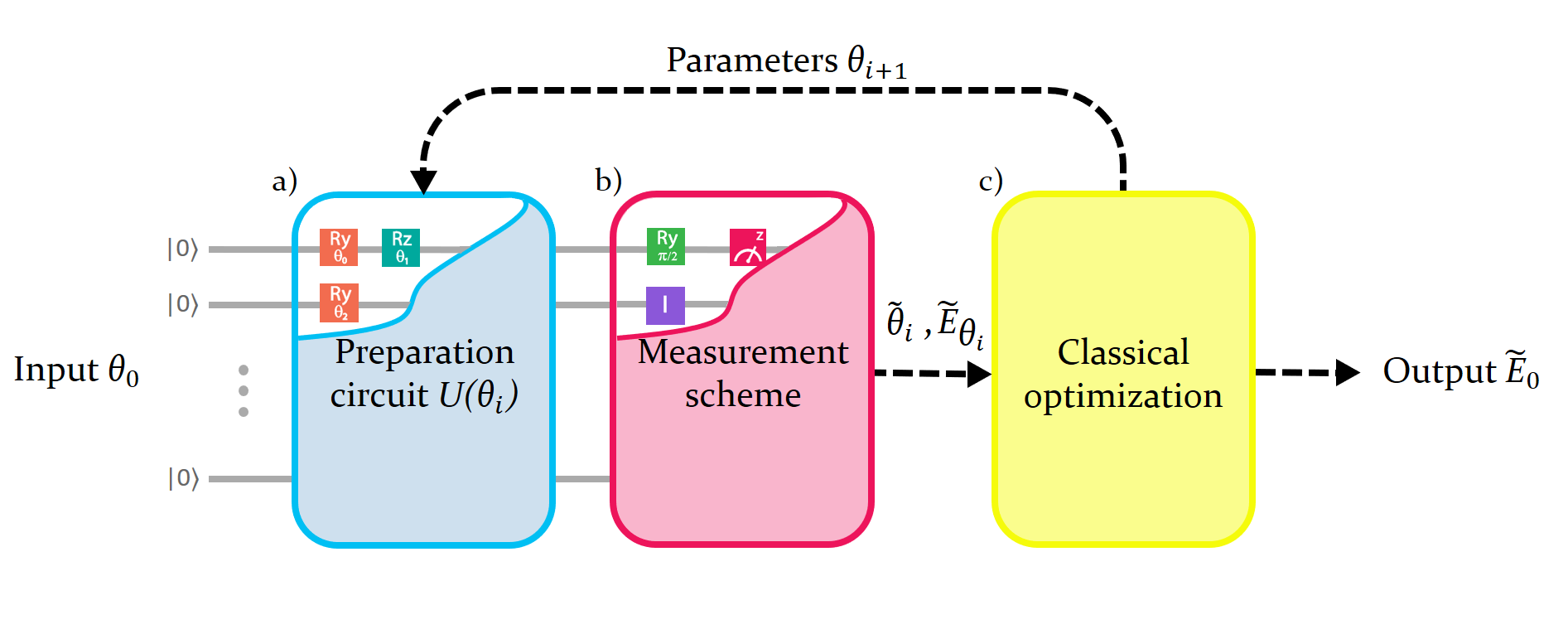}
    \caption{Flow chart of VQE. The parameters are encoded into rotation gates to prepare the trail quantum state. Then it is measured in different basis using a quantum computer to get an estimate of the expectation of the Hamiltonian. Based on the output cost function value, classical optimizer modifies the parameters to give a new trail state. This process is iteratively repeated to obtain the ground state.}
    \label{fig:VQE}
\end{figure}

This energy estimate serves as a cost function that is minimized using a classical optimizer, which iteratively updates the parameters $\boldsymbol{\theta}$ in order to approach the system’s ground state.
The optimization loop continues until convergence is achieved, ideally resulting in a parameterized state that approximates the true ground state of the molecule or system under study. A pictorial representation of the VQE is shown in Fig. \ref{fig:VQE}.

In recent years, VQE has become a standard benchmark for variational quantum algorithms, not only in quantum chemistry but also in condensed matter physics and optimization~\cite{cerezo2021variational,tilly2022variational}.
However, its performance is highly sensitive to the structure of the ansatz, the efficiency of the classical optimizer, and the effects of quantum noise. \\

\newpage
\subsection{Challenges for Optimization in VQE} \label{sec:vqe-challenges}

While the general structure of VQE appears straightforward, several aspects of its optimization landscape pose significant challenges.
In particular, the algorithm is often hindered by three key factors: barren plateaus, periodic cost functions, and stochastic noise arising from quantum measurements.

Barren plateaus refer to regions in the parameter landscape of variational quantum algorithms (VQAs) where the gradient of the cost function becomes exponentially small as the system size increases~\cite{mcclean2018barren}. In these regions, the optimization landscape is essentially flat, providing negligible information about how to adjust parameters to lower the cost function. This phenomenon severely hampers the performance of gradient-based optimization methods, which rely on the magnitude and direction of gradients to iteratively improve the parameters. The problem becomes particularly pronounced in deep quantum circuits or when using highly expressive ansatzes that span a large portion of the Hilbert space. In such cases, the random initialization of parameters typically places the system in a barren plateau. This significantly limits the scalability of variational quantum algorithms, as increasing the number of qubits or circuit depth often exacerbates the issue, rendering large-scale implementations ineffective unless specific strategies are employed to mitigate the barren plateau problem.

Periodic cost functions naturally arise in variational quantum computing due to the structure of parameterized quantum gates—particularly rotational gates such as $R_x(\theta)$, $R_y(\theta)$ and $R_z(\theta)$ -- whose behavior is inherently periodic with respect to the parameters. While this cyclic topology is well-defined within the formalism of quantum mechanics, classical optimization algorithms are often designed under Euclidean assumptions of linear, unbounded parameter spaces. 

\begin{figure}
    \centering
    \includegraphics[width=0.5\linewidth]{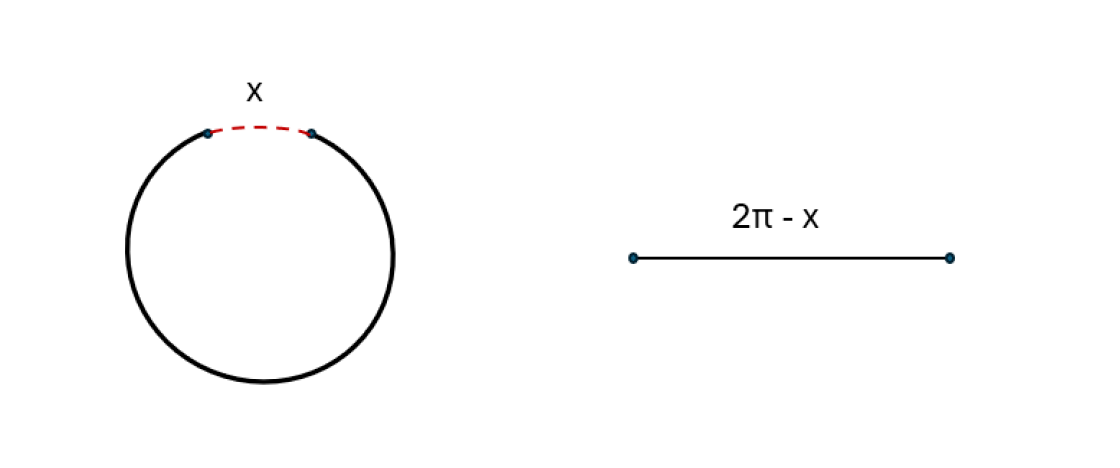}
    \caption{Illustration of distance measurement in periodic (left) and non-periodic(right) spaces. In the periodic case, the shortest distance between two points on the circle is along the arc of length $x$. In the non-periodic (linear) case, the distance between the same two points is measured as $2\pi - x$.}
    \label{fig:periodicity}
\end{figure}

This mismatch introduces a fundamental incompatibility between the topology of the quantum cost landscape and the assumptions made by many traditional classical optimizers. 
Specifically, classical routines may misinterpret two points on a circular manifold (e.g., $\theta = 0  ~\&~ \theta =2\pi$) as being maximally distant, when they are in fact adjacent in the periodic topology, as shown in Fig. \ref{fig:periodicity}. 
This misalignment can lead to several practical issues: inefficient exploration of the parameter space, distorted gradient directions near the boundaries, and incorrect detection of local minima.
Similarly, evolutionary algorithms that rely on Euclidean distance may overestimate differences between points, degrading performance. 
These challenges are further amplified in high-dimensional settings such as VQE, where multiple parameters are periodic.
Commonly used strategies to circumvent this issues by creating boundaries deem to be ineffective.
Hard boundaries create artificial discontinuities at the edges of the parameter space, disrupting smooth optimization e.g., near $0$ and $2\pi$.
Soft boundaries attempt to smooth these edges but can introduce distortions or local minima.
Both approaches fail to respect the true circular topology, often leading to inefficient exploration and suboptimal convergence. 

Though an obvious aspect of quantum computation, measurement noise remains a significant challenge for classical optimization methods. It introduces stochasticity into cost function evaluations, as quantum expectation values are estimated from a finite number of measurements (shots). This noise can obscure genuine trends in the cost landscape, destabilize the optimizer’s trajectory, and amplify parameter noise, particularly in already fragile regions. This effect is especially prone for gradient based methods where a small difference in the value of cost function can significantly harm the direction of the locally computed gradient. 

Together, these factors expose the limitations of many standard optimization techniques in the quantum setting~\cite{cerezo2021challenges}.
In particular, methods that do not account for noise, flat regions, or the periodic nature of the parameter space can struggle to converge reliably or may get stuck in a local minima—especially as the dimensionality of the system increases.

\section{HOPSO and its modifications}\label{sec: method}
The Harmonic Oscillator-based Particle Swarm Optimization (HOPSO) algorithm~\cite{10.1371/journal.pone.0326173} was introduced as a physically motivated alternative to conventional Particle Swarm Optimization (PSO) variants.
Inspired by the dynamics of damped harmonic oscillators, HOPSO models each particle’s trajectory in the search space using time-evolving sinusoidal motion around an attractor, governed by decaying amplitude and phase-shifted oscillations. 
This can be visualized as each particle being connected to a damped spring anchored at the attractor.
This mechanism encourages broad initial exploration followed by gradual convergence, mimicking natural stabilization behavior. To get a better understanding of HOPSO, refer to appendix \ref{A1-HOPSO}, or \cite{10.1371/journal.pone.0326173} for more detailed explanation.


In our prior work~\cite{10.1371/journal.pone.0326173}, HOPSO demonstrated strong performance across a range of classical benchmark functions including Rastrigin, Ackley, Griewank, and the Rosenbrock function, outperforming several widely used metaheuristics in both low- and high-dimensional settings. Its structured dynamics, minimal reliance on randomness, and physical interpretability make it a promising candidate for adaptation to quantum optimization problems, where noisy cost evaluations and periodic parameter domains often create problems for traditional methods. 

\subsection{Inactive Particle Removal} \label{sec:dead-particles}

While HOPSO seems to be a good candidate for utilization in VQE, we suggest a few modifications that enhance its working. The first one is a general update that is not tailored to quantum-related problems. 

During the course of optimization, numerical instabilities can arise--particularly when calculating the phase $\theta$ via the $arccos$ function \eqref{theta_eq}. This occurs if $\theta$ falls outside the mathematically defined domain due to round-off errors or invalid position updates.

A common fix is to clip the output of $\cos \theta$ to the range $[-1, 1]$, but this introduces discontinuities and may break the harmonic oscillator's smooth behavior. Instead, we adopt a stricter policy: when a particle reaches a state where its phase becomes undefined or invalid, it is marked as inactive and removed from further iterations.

This decision is justified by the fact that such a particle is unlikely to contribute meaningfully to exploration, especially in later optimization stages when global convergence is approaching. Terminating these particles improves overall efficiency by avoiding wasted resources on unproductive trajectories, and helps maintain coherent swarm dynamics.

\subsection{Modifications accounting for periodic landscape}
When HOPSO is applied to optimize quantum landscapes via the Variational Quantum Eigensolver (VQE), modifications are necessary to address for the specific structure of quantum parameter landscapes (periodicity).

Quantum cost functions differ from classical ones in two fundamental ways: they are periodic (with a period of $2\pi$) and are evaluated stochastically due to quantum measurement noise. Standard HOPSO does not explicitly account for periodic boundaries, and its unmodified version may produce attractors or updates that lie outside the valid domain of parameters. Additionally, fluctuating cost evaluations due to finite sampling (shot noise) can introduce instability in the optimization process.

As discussed earlier in Section~\ref{sec:vqe-challenges}, both soft and hard boundary conditions fail to fully address this problem.
To mitigate it, we apply a very specific strategy tailored for HOPSO optimizer, being a combination of free movement and restricted optimization. Namely, we let every particle move freely in the parameter space, not being restricted by any kind of boundaries. However, if a best-position is found, contributing to form the attractor for that (personal best) or all (global best) particles, that position is clipped to a specified region (if they are identified outside that region, a $2\pi$ modulo is used to shift them into the desired region). 

This procedure combines the advantages of the free unrestricted movement of particles with the fact that they are attracted only to a well constrained region. As a result, they will smoothly converge to that region, while maintaining the original shape of the cost function. 

It would seem natural to select a fixed interval such as \([0, 2\pi)\) for the allowed domain of best positions. However, if the optimal value of the parameter is close to the selected boundary ($0$ in this case), oscillations can be introduced due to periodic change of the attractor between $\epsilon$ and $2\pi - \epsilon$. To avoid this problems, we introduce randomness into the choice of domain by shifting the range to \([r, r + 2\pi)\), where \(r\) is selected randomly between $0$ and $2\pi$.

Keeping this in mind, to adapt HOPSO for use in the VQE framework, we introduce three key modifications. 




\subsubsection{Modulo modification of Best Positions} \label{sec:modulo-correction}
As described above, we apply the modulo correction when a new personal best position is identified and found to fall outside the chosen angular range $[r,r+2\pi)$.
In such cases, the personal best position is relocated to an equivalent point within $[r, r + 2\pi)$ using modular arithmetic. 
It should be noted that the particle's position remains untouched. 
The attractor is then recalculated accordingly. This localized correction preserves the swarming behavior around the attractor while still respecting the periodic geometry of the parameter space.

The same principle is applied to the global best, which is always selected from among the updated personal bests.
This ensures stable, periodic-aware search behavior while minimizing overhead.

Having confined best positions to a consistent angular window via selective modulo updates, we now specify how to compute the attractor on the circle so that weighted averages respect wrap-around.

\subsubsection{Attractor Computation} \label{sec:attractor-correction}

If one simply takes a linear combination of a particle’s personal best $p$ and the global best $g$, the resulting attractor can end up far from both points—even though, by wrapping around the boundary, there exists an equivalent attractor on the opposite side that is much closer. To mitigate this problem, we do the following:

\begin{itemize}
    \item If $|p - g| < \pi$, the attractor \eqref{attractor} is computed as a standard weighted average:
    \[
    a = \frac{c_0 p + c_1 g}{c_0 + c_1}
    \]
    
    \item If $|p - g| > \pi$, the shortest distance between $p$ and $g$ lies around the periodic boundary rather than across the linear interval.
    In such cases, a naive weighted average computed linearly will fall in the wrong direction — effectively averaging across the longer arc. To address this, we apply a coordinate shift and use modulo arithmetic to compute the correct average along the circular topology:
    \[
    a = \left( \frac{c_0 p + c_1 g}{c_0 + c_1} + \frac{2\pi}{c_0+c_1} - r \right) \bmod 2\pi + r
    \]
    where $r$ is a uniformly sampled reference offset from $[0, 2\pi)$, selected at the beginning of the optimization run to define the valid angular window.
\end{itemize}

Building on the periodic attractor, we define a per-dimension amplitude floor proportional to the wrapped 
$|p-g|$ distance so that oscillations remain meaningful near the boundary.

\subsubsection{Threshold Amplitude}
Last but not the least, once the attractor is computed, we must calculate the wrapped distance between the personal best $p$ and the global best $g$ to ensure that the amplitude decays only to a meaningful level \eqref{th_A}. We first compute the wrapped angular difference:

\[
\text{remainder} = |p_{j,d} - g_d| \bmod 2\pi
\]
The effective amplitude threshold $A_{\text{th},j,d}$ is computed by choosing the minimum distance between the personal and global best positions:
\[
A_{\text{th},j,d} = \left( \frac{\min(2\pi - \text{remainder},\ \text{remainder})}{2} \right) \cdot m,
\]
where $m$ is the multiplier for amplitude decay, typically $m = 2.05$ as used in our experiments.






\vspace{1.5em}
\noindent
The pseudocode for periodic HOPSO (taking into account these modifications) is presented below. 

\vspace{1em}

\begin{algorithm}[H]
\caption{HOPSO: Harmonic Oscillator-based PSO for VQE}
\label{alg:hopso}
\begin{algorithmic}[1]
\Require Cost function \texttt{cost\_fn}, hyperparameters \texttt{hp}, number of particles \texttt{num\_particles}, dimension, max iterations, runs
\Ensure Minimum energy per run $e_{\min}$
\For{\textbf{each} run \textbf{do}}
    \State Initialize random reference shift $r \sim \mathcal{U}[0, 2\pi)$;
    \State Initialize particle positions and velocities;
    \State Evaluate cost and set personal bests;
    \State Set initial global best from personal bests;
    \State Compute initial attractors, amplitudes, and phases $\theta$;
    \For{\textbf{each} iteration}
        \For{\textbf{each} particle}
            \If{particle is \textbf{not} dead}
                \State Update time $t$ and decay amplitude $A$;
                \State Enforce minimum amplitude constraint;
                \State Update position and velocity;
                \State Evaluate cost at new position;
                \If{\textit{new cost improves personal best}}
                    \State Update personal best (modulo-corrected);
                    \State Recompute attractor, amplitude, and phase $\theta$;
                    \If{\textit{invalid} $\theta$ (\textit{out of domain})}
                        \State Mark particle as dead;
                    \EndIf
                \EndIf
            \EndIf
        \EndFor
        \State Check and update global best across swarm;
        \If{\textit{global best changed}}
            \State Reset time $t$ for all particles;
            \State Recompute swarm-wide attractors, amplitudes, and $\theta$;
            \State Mark invalid particles as dead;
        \EndIf
    \EndFor
    \State Store final results of this run;
\EndFor
\end{algorithmic}
\end{algorithm}

\section{Results on \texorpdfstring{H\textsubscript{2}}{H2}}\label{sec4}

To evaluate the proposed algorithm in a quantum context, we first consider the hydrogen molecule (H\textsubscript{2}), represented using a 4-qubit Hamiltonian derived via a Jordan–Wigner transformation. This yields a weighted sum of Pauli strings \cite{mihalikova2022best}:

\begin{equation}
\begin{aligned}
H =\ & -0.80718\,IIII + 0.17374\,ZIII - 0.23047\,ZZII + 0.17374\,IIZI \\
     & - 0.23047\,IZZZ + 0.12149\,IZII + 0.16940\,IZZI - 0.04509\,ZXXI \\
     & + 0.04509\,XIXZ + 0.04509\,XIXI - 0.04509\,XZXZ + 0.16658\,ZZZZ \\
     & + 0.16658\,ZZZI + 0.12149\,IZIZ.
\end{aligned}
\end{equation}

To approximate the ground state, we employ a hardware-efficient ansatz \cite{kandala2017hardware} with three layers of parameterized single-qubit rotations ($R_y$ and $R_z$ gates), sandwiched between entangling CNOT operations arranged in a linear chain topology. The total number of parameters is 32. This design balances expressibility with low circuit depth, ensuring feasibility in near-term devices. A visual of this ansatz is shown below, see Fig.~\ref{fig:h2_ansatz}.

\begin{figure}[ht]
    \centering
    \includegraphics[width=\linewidth]{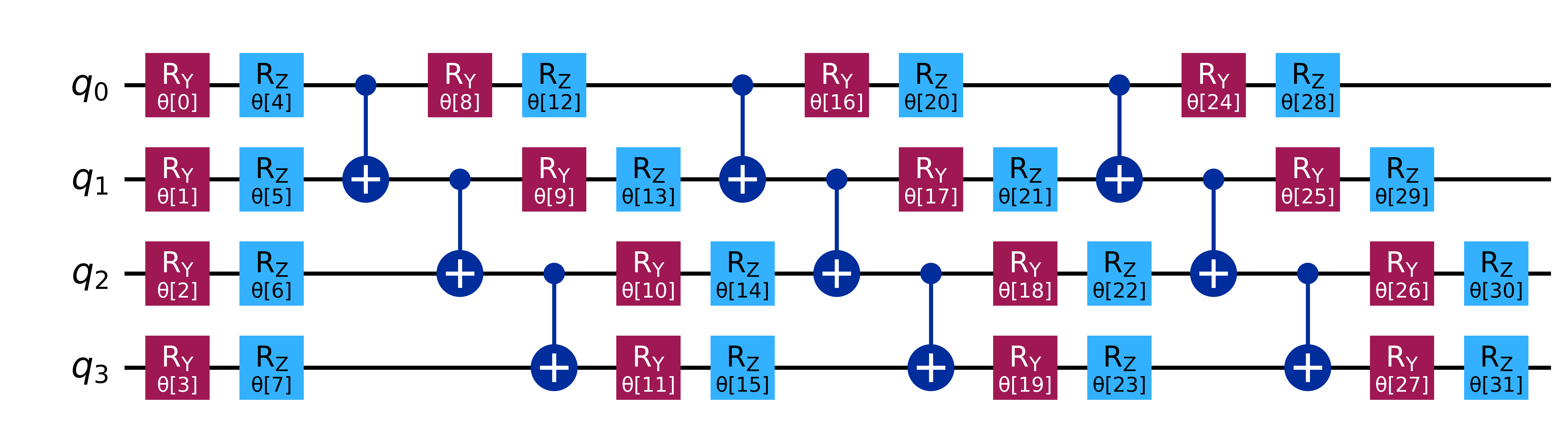}
    \caption{Three-layer hardware-efficient ansatz with alternating $R_y$/$R_z$ rotations and linear entanglement structure.}
    \label{fig:h2_ansatz}
\end{figure}

\subsection{Performance in the Absence of Noise} \label{sec:h2-noiseless}

We assessed the performance of the periodic HOPSO algorithm by comparing it with three widely used optimizers: standard PSO, COBYLA, and DE. All experiments were conducted under noiseless conditions using a total budget of 5000 cost function evaluations.

For the population-based methods (HOPSO and PSO), we used 10 particles over 500 iterations. HOPSO was configured with $\lambda = 0.1$, $c_1 = c_2 = 1$, and $m = 2.05$. The standard PSO baseline used $c_1 = c_2 = 2.05$ (attraction weights towards personal and global best positions) with $\chi = 0.729$ (constriction factor)~\cite{clerc}. COBYLA was run with its default settings and capped at 5000 iterations. DE used a population size of 32 (matching problem dimensionality), and its iteration count was set to 157 to fit the same overall budget.

All optimizations were implemented in Qiskit 2.0.0, and cost evaluations were performed using the Statevector-Estimator module to ensure deterministic outcomes. COBYLA, DE were implemented using $Scipy$ module ~\cite{2020SciPy-NMeth} in python, where as HOPSO and PSO were implemented in-house.

\begin{figure}[ht]
    \centering
    \includegraphics[width=0.85\linewidth]{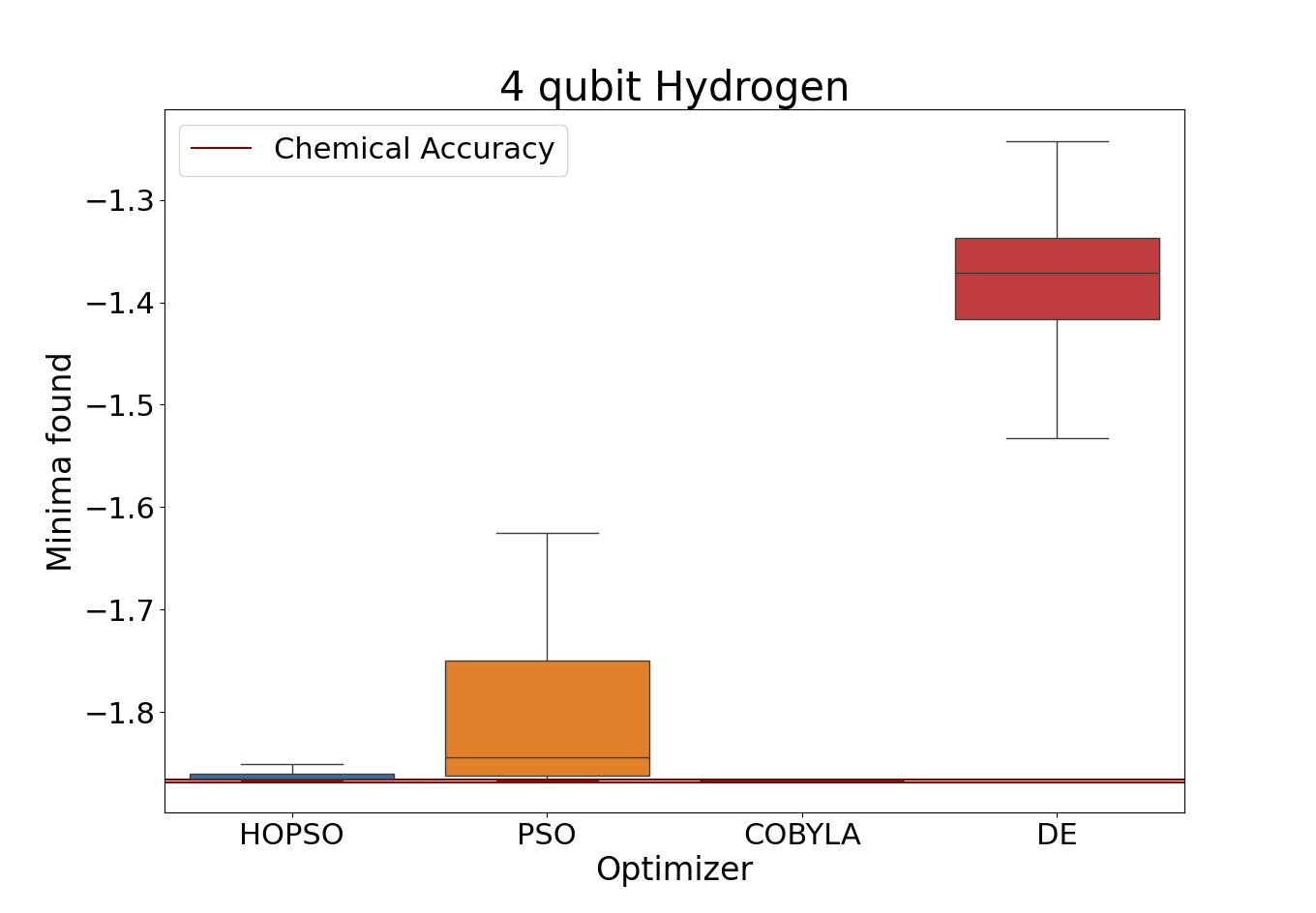}
    \caption{Minimum energies found by each optimizer for 4-qubit H\textsubscript{2} system without noise. Red line indicates chemical accuracy threshold.}
    \label{fig:h2_boxplot_noiseless}
\end{figure}

The comparison of results is summarized in Fig.~\ref{fig:h2_boxplot_noiseless}. Surprisingly, COBYLA outperformed the other methods in this setting, despite often lagging behind on classical benchmarks \cite{10.1371/journal.pone.0326173}. In contrast, DE, which typically performs well classically, showed poor convergence in the quantum case.

This reversal can be attributed to differences in the structure of quantum landscapes. In particular, the periodicity and smoothness of cost landscapes in quantum optimization appear well-suited to COBYLA's local, linear approximation strategy. DE, on the other hand, lacked the robustness needed under a tight budget without its typical post-processing refinement step (e.g., polish=True in $SciPy$).

\subsection{Performance under Shot Noise} \label{sec:h2-noise}

To evaluate each optimizer’s robustness to realistic quantum noise, we repeated the previous experiments with simulated shot noise. Specifically, we set the number of shots to 1000 in Qiskit's estimation module, introducing statistical fluctuations into the cost evaluations.

\begin{figure}[ht]
    \centering
    \includegraphics[width=0.85\linewidth]{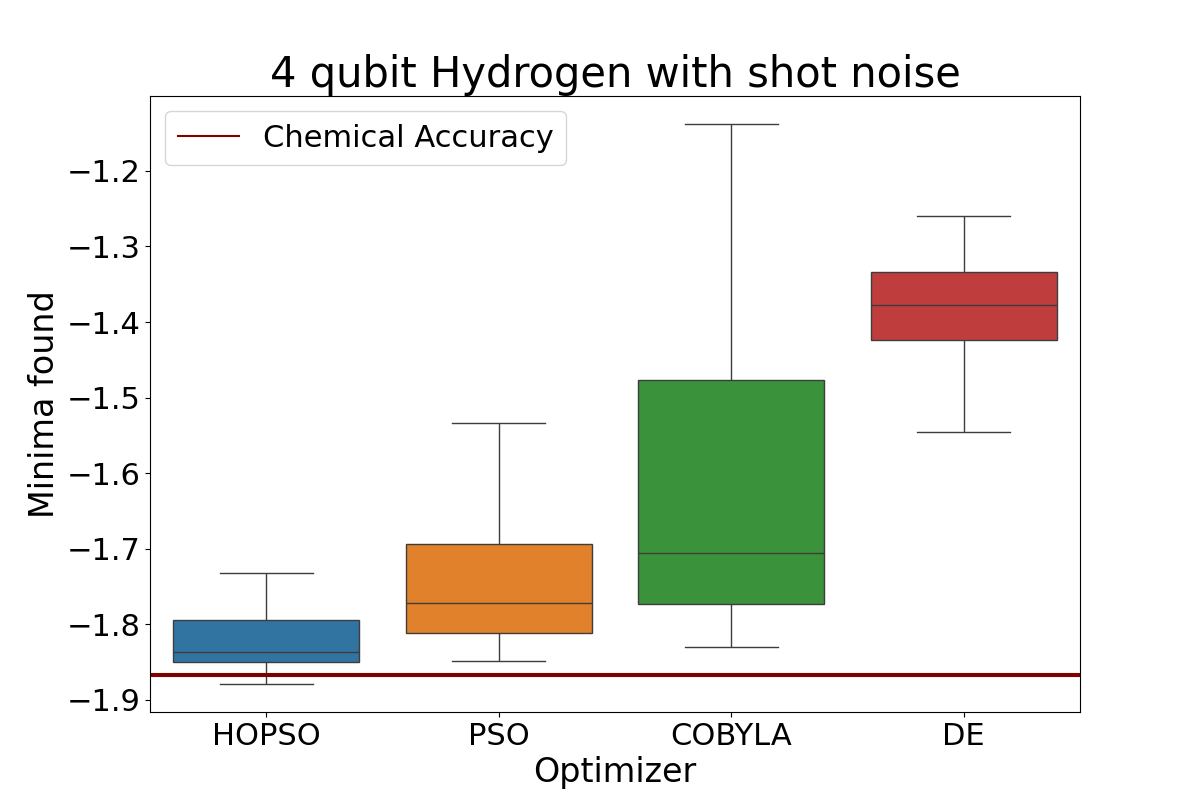}
    \caption{Minimum energies found by each optimizer for the 4-qubit H\textsubscript{2} system with \texttt{shots=1000}. Note that the energies shown in the figure are the measured energies that are the outputs of VQE optimization. This is why a fraction of the energies can, due to shot noise, also gain values below the real exact ground state energy. Red line marks chemical accuracy threshold.}
    \label{fig:h2_boxplot_noise}
\end{figure}

The results, summarized in Fig.~\ref{fig:h2_boxplot_noise}, reveal significant shifts in optimizer performance. COBYLA, which had previously excelled in the noiseless setting, now struggled to maintain stability and accuracy. Its reliance on smooth local approximations made it especially vulnerable to fluctuations in the cost function caused by sampling noise.

In contrast, population-based optimizers—HOPSO, PSO, and DE—proved more resilient. These methods inherently evaluate multiple candidate solutions in parallel, which helps average out noise across the swarm. Among them, HOPSO demonstrated the best overall performance. Its oscillatory dynamics contributed to reliable exploration despite stochastic measurements.

These findings underscore the importance of using noise-tolerant strategies in near-term quantum optimization. In environments where measurement noise is unavoidable, optimizers that can integrate noisy feedback across diverse particles are better equipped to reach consistent solutions. The modification to account for the periodicity of quantum landscapes has enabled HOPSO to outperform other methods in noisy environments.

\section{Performance with LiH} \label{sec5}

We next evaluate the optimizers on a more complex quantum system: the Lithium Hydride (LiH) molecule. The LiH Hamiltonian was constructed using Qiskit Nature’s electronic structure tools (PySCFDriver~\cite{PySCF} which calculated the molecular integrals representing the electronic Hamiltonian), with geometry defined by placing lithium at the origin and hydrogen at a bond distance of 1.5474~\AA. The STO-3G basis set was used for the underlying quantum chemistry calculations, and the electronic Hamiltonian was transformed to a qubit Hamiltonian via the Jordan-Wigner mapping.

To reduce computational cost, we applied symmetry tapering to the full 12-qubit Hamiltonian~\cite{bravyi2017taperingqubitssimulatefermionic}, yielding an 8-qubit effective representation while preserving key energy characteristics. 

The resulting 8-qubit Hamiltonian is given by:
\begin{align}
H_{\text{LiH}} = &\ -4.98851\, I I I I I I I Z - 0.11677\, I I I I I Z I I + 1.00871\, Z Z I I Z Z Z Z \nonumber \\
&+ 0.08981\, Z Z I I Z Z I I - 0.00761\, I I I I I I Y Y + 0.00022\, Z Z I I Z X Z X \nonumber \\
&- 0.00761\, I I I I I I X X + \cdots
\label{eq:lih-hamiltonian}
\end{align}
This Hamiltonian includes over 200 Pauli strings, but only the terms with the highest weights are shown here for brevity. 

Exact diagonalization of the resulting matrix confirmed a ground-state energy near $-8.908697$ Ha. This procedure is  a standard benchmark for quantum compuational chemistry algorithms due to its balance between complexity and tractability.

\vspace{1em}

\begin{figure}[ht]
    \centering
    \includegraphics[width=0.85\linewidth]{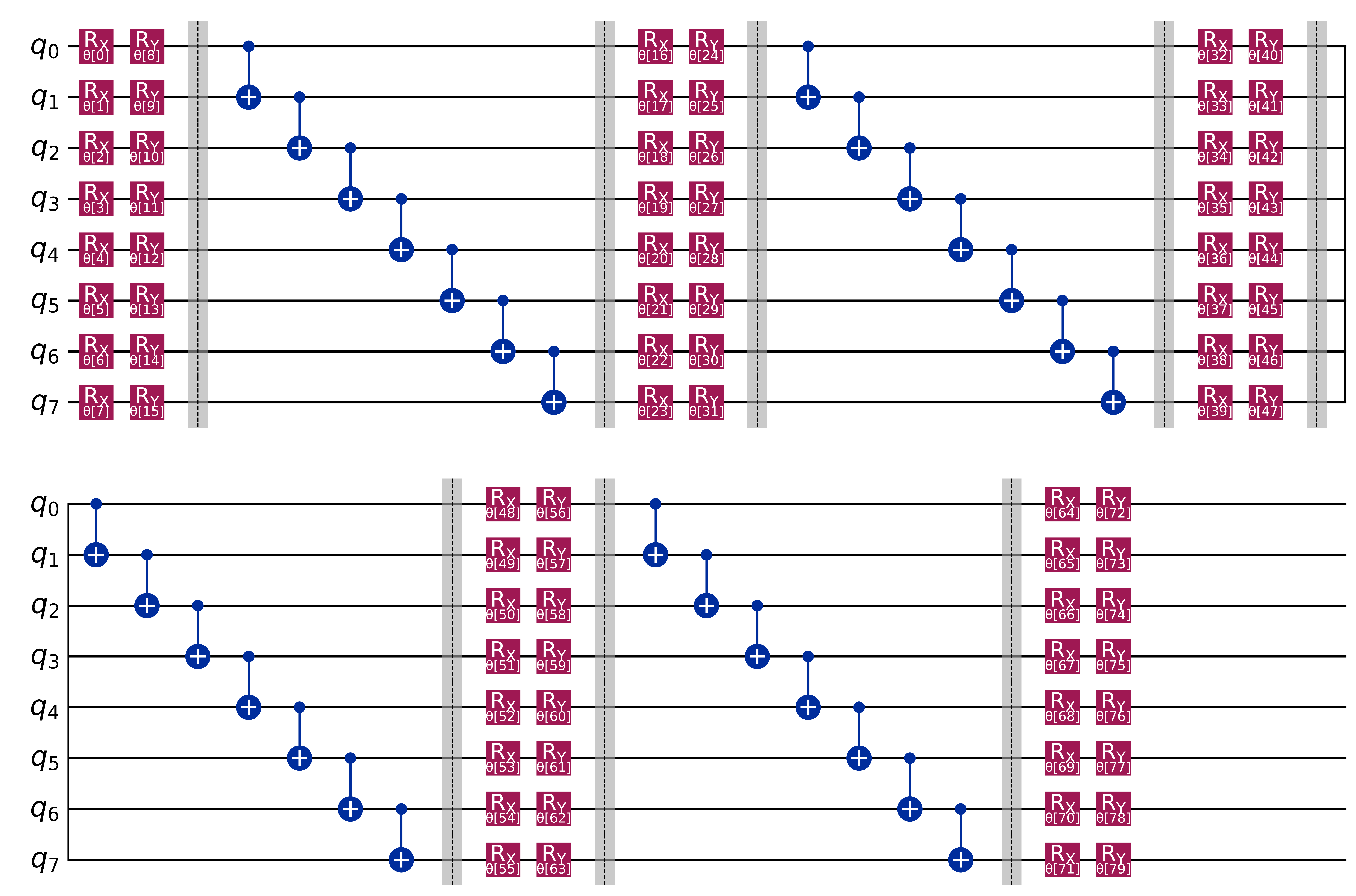}
    \caption{Hardware-efficient 8-qubit ansatz used for LiH simulations, with alternating single-qubit rotations and staircase entanglement. 4 layer-repetitions containing 80 parameters with barriers between layers for clarity.}
    \label{fig:lih_ansatz}
\end{figure}

\noindent\textbf{Ansatz structure.} 
To tackle this high-dimensional optimization space (80 parameters), we used a hardware-efficient variational 8-qubit ansatz built with Qiskit. The circuit alternates between $R_y$ and $R_z$ gates on each qubit, with four layers (reps = 4) of linear entanglement operations through CNOTs connecting adjacent qubits. These hardware-efficient circuits balance expressivity and
implementability. Barriers were inserted between repetitions to separate the layers, see Fig.~\ref{fig:lih_ansatz}.

\noindent\textbf{Hardware and resource considerations.} Due to the increased complexity of the 8-qubit ansatz - comprising 80 tunable parameters (an 80-dimensional optimization landscape) - we employed a highly parallel setup.
Experiments were carried out on the Karolina~\cite{karolina2021} and Devana~\cite{devana} supercomputers, using 400 CPU cores distributed in multiple nodes to run global optimization algorithms in parallel.
Each optimizer was allowed 800,000 function evaluations per run, corresponding to 10,000 iterations over 80  particles. COBYLA was excluded because of its inherently sequential structure, and therefore, it was not easily possible to parallelize.
This omission allowed for more efficient exploration of the parameter space using population-based methods, which are well suited to the parallel resources available.
The estimated core-hour budget for the entire LiH study exceeded 50,000, demonstrating the extensive amount of resources required for even relatively small molecular systems using variational quantum algorithms.

\vspace{1em}

\subsection{Performance in noiseless scenario}

For HOPSO, we used the following settings: $\lambda = 0.008$, $c_1 = c_2 = 1$, and multiplier $m = 2.05$, with the upper time limit set to $t_{\text{ul}} = 2\pi$.
The optimizer was allowed to run for 10,000 iterations per particle, totaling 800,000 function evaluations per run. 
PSO used the same number of particles and iterations, with $c_1 = c_2 = 2.05$ and constriction factor $\chi = 0.729$.
For Differential Evolution (DE), we used a population size of 80 and set the maximum iteration count to 10,000 to match the computational budget.

To ensure consistent comparison, DE was run using the `deferred' update strategy from SciPy, which evaluates the entire population in batch before generating new candidates—well suited for parallel evaluation. This helped avoid excessive overhead from sequential updates common in `immediate' mode.

\begin{figure}[ht]
    \centering
    \includegraphics[width=0.85\linewidth]{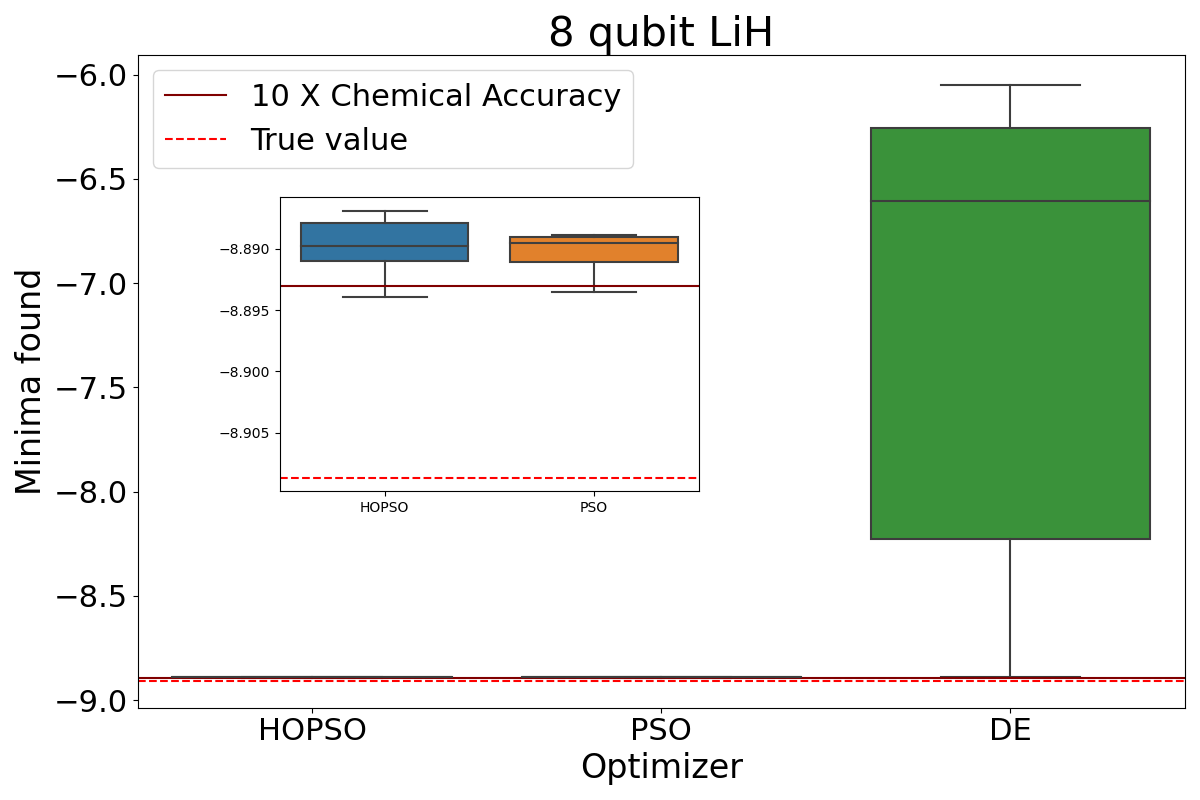}
    \caption{Minimum energies found by each optimizer for the 8-qubit LiH system using \texttt{statevector simulator}. Red line marks chemical accuracy threshold.}
    \label{fig:lih_noiseless_box}
\end{figure}

Fig.~\ref{fig:lih_noiseless_box} presents the distribution of the minimum energies found by each optimizer. HOPSO and PSO both achieved tight clusters near the known ground state energy of LiH, with HOPSO exhibiting marginally higher accuracy and stability. In contrast, DE showed a broader spread of outcomes and frequent deviations from the chemical accuracy threshold.

These results reinforce HOPSO’s strength in high-dimensional problems where parallel resources are available. Its structured dynamics contribute to good convergence performance. PSO also performed well, though with slightly more variance. DE, despite its success on classical benchmark functions, showed limitations in this quantum setting.

\subsection{Performance in the Presence of Shot Noise}

To evaluate the robustness of the optimizers under realistic quantum conditions, we repeated the LiH simulations with shot noise enabled.
Shot noise was introduced using Qiskit's Aer simulator by setting the number of measurement shots to 1000.
This simulates the stochastic fluctuations arising from finite sampling during quantum measurements, a common challenge in real quantum hardware. The optimizer settings were kept identical to the noiseless case.

\begin{figure}[ht]
    \centering
    \includegraphics[width=0.85\linewidth]{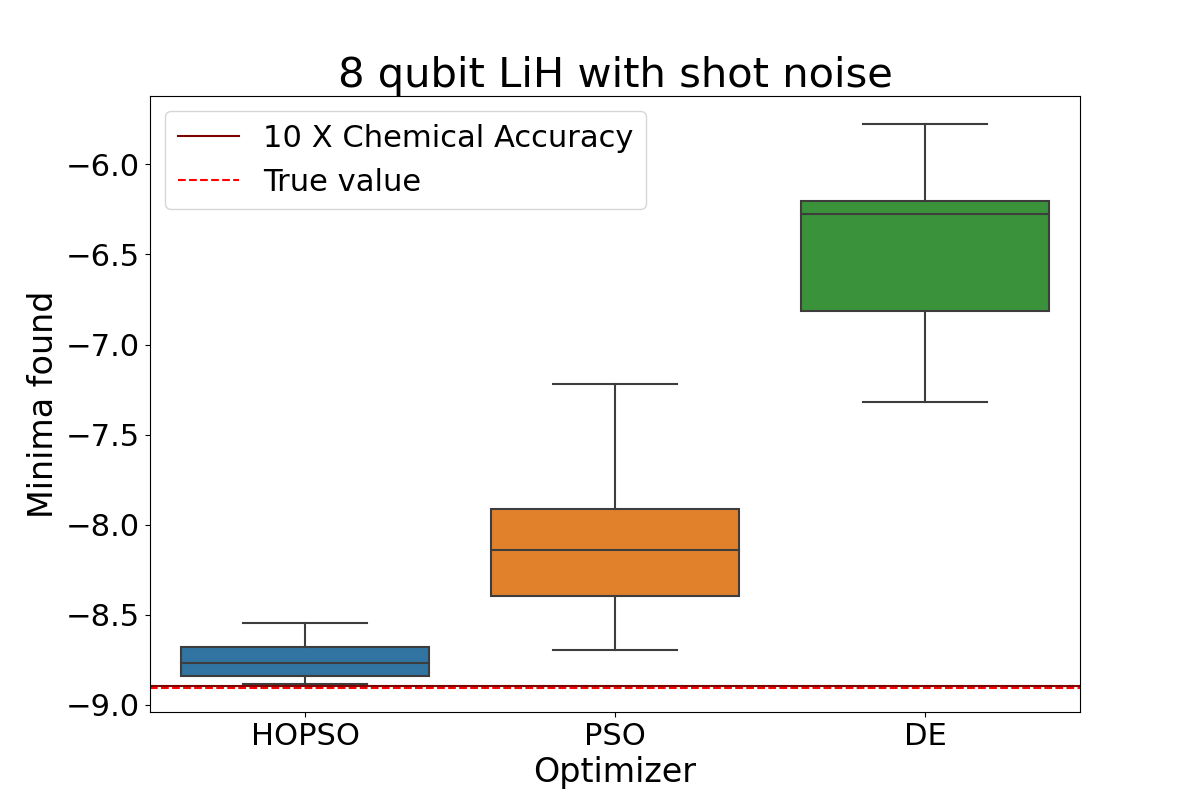}
    \caption{Minimum energies found by each optimizer for the 8-qubit LiH system with \texttt{shots=1000}. Note that the energies shown here are measured energies. Red line marks chemical accuracy threshold.}
    \label{fig:lih_noisy}
\end{figure}

\vspace{1em}

The results, shown in Fig.~\ref{fig:lih_noisy}, demonstrate clear differences in optimizer robustness. HOPSO maintained strong performance despite the presence of noise, consistently producing energy estimates close to the true ground state with minimal variance.
PSO, while still competitive, displayed a wider spread of energy values, indicating increased sensitivity to measurement fluctuations.
DE exhibited the poorest performance under noise, producing less accurate results with substantial variance.

These findings underscore the value of population-based optimizers—such as HOPSO, PSO, and DE—in a quantum search space. However, despite also being population-based, DE’s differential mutation strategy proved less robust in navigating the high-dimensional, noisy landscape of the LiH problem compared to the oscillator-guided dynamics of HOPSO and PSO.

Taken together, these results support the value of algorithmic robustness in VQE optimization under noisy conditions. In scenarios where noise and measurement cost are non-negligible, optimizers that have population dynamics—such as HOPSO—offer a promising path toward achieving reliable quantum solutions under constrained resource budgets.

\section{Conclusion}
\label{sec:conclusion}

This work introduces and evaluates a modified version of the Harmonic Oscillator Particle Swarm Optimization (HOPSO) algorithm as a classical optimizer within the Variational Quantum Eigensolver (VQE) framework. While our previous publication~\cite{10.1371/journal.pone.0326173} introduced the algorithm and its performance strictly comparing to other classical optimization methods, this paper builds on those results by applying this algorithm to a quantum setting, and compares how its abilities to the other classical optimizers specifically within this context. 

Our results suggest that the structured, physically-inspired dynamics of HOPSO are well-suited for navigating the complex, \textit{periodic}, and noisy landscapes characteristic of VQE. In particular, we show that HOPSO's built-in mechanisms for attractor adjustment and amplitude damping contribute to its robustness against local minima and cost-function noise, while enabling efficient exploration and exploitation trade-offs.

These findings support the broader argument that optimization in quantum-classical hybrid algorithms may benefit from revisiting classical methods grounded in physical principles, rather than relying solely on black-box heuristics or gradient descent methods. Future work will explore adaptive versions of HOPSO, hybrid quantum-classical co-design strategies, and its integration with error mitigation and ansatz adaptation protocols on real quantum hardware. 

\bmhead{Acknowledgements}
The authors thank Martin Friak and his team for valuable discussions. This work was supported by the VEGA project No. 2/0055/23. MP and IAM acknowledge support from the Research and Innovation Authority projects 09I03-03-V04-00425 and 09I03-03-V04-00685. YC acknowledges support from the SAS Doktogrant APP0683. The funders had no role in study design, data collection and analysis, decision to publish, or preparation of the manuscript.

This work was also supported by the Ministry of Education, Youth and Sports of the Czech Republic through e-INFRA CZ (ID: 90254) for use of the Karolina supercomputer. 
Part of the research results was obtained using the computational resources procured in the national project National competence centre, Devana for high performance computing (project code: 311070AKF2) funded by European Regional Development Fund, EU Structural Funds Informatization of society, Operational Program Integrated Infrastructure.

\bibliography{sn-bibliography}

\begin{appendices}

\section{General working of HOPSO}\label{A1-HOPSO}
Unlike standard PSO—where particles can exhibit sudden velocity shifts—our framework models each particle as a harmonic oscillator. Mechanical potential energy accumulates whenever a particle moves away from its designated attractor and then converts into kinetic energy as it returns, with the sum of these two energies conserved aside from any applied damping. Importantly, this mechanical‐energy bookkeeping is entirely separate from the algorithm’s objective function, so particles follow smooth, stable trajectories without abrupt velocity discontinuities.

When modeling a single spring in multiple dimensions, the total energy becomes a complex, non-trivial function of the particle’s position.
To simplify this, we propose assigning an independent virtual spring to each dimension, attracting the particle only along that axis.
This allows for straightforward energy calculations and enables constraints in each dimension to be handled independently.
Since modeling the full continuous motion of the particle is unnecessary, we instead take discrete snapshots of its position at randomly selected time points.
This random sampling introduces stochasticity into the optimization process.

\subsubsection{General working of HOPSO}

Each particle's position in a given dimension follows the trajectory of an underdamped harmonic oscillator:
\begin{equation}
    x(t) = A_0 e^{-\lambda t} \cos(\omega t + \theta) + x_0,
\end{equation}
where \( A_0 \), \( \lambda \), \( \omega \), and \( \theta \) are the initial amplitude, damping factor, angular frequency, and phase, respectively, and \( x_0 \) is the attractor's position. The velocity is given by:
\begin{equation}
    v(t) = -\omega A_0 e^{-\lambda t} \sin(\omega t + \theta) - \lambda A_0 e^{-\lambda t} \cos(\omega t + \theta).
\end{equation}

To simplify multidimensional dynamics, we model each dimension with an independent spring. Particles' positions are sampled at random time points \( t \in [0, t_{ul}] \), introducing stochasticity. For full oscillation coverage, we set \( t_{ul} = 2\pi \), and time is updated as:
\begin{equation}
    t_{i+1} = t_i + \text{rand}[0, t_{ul}].
\end{equation}

Particles are initialized with random positions and velocities. Each particle oscillates around an attractor defined as:
\begin{equation} \label{attractor}
    a_{j,d} = \frac{c_1 p_{j,d} + c_2 g_d}{c_1 + c_2},
\end{equation}
where \( p_{j,d} \) and \( g_d \) are the personal and global best positions in dimension \( d \). Typically, \( c_1 = c_2 \), placing the attractor midway.

The amplitude and phase are calculated at \( t=0 \) as:
\begin{equation}
    A_0 = \sqrt{(x(0)-a)^2 + \frac{(v(0) + \lambda(x(0) - a))^2}{\omega^2}},
\end{equation}
\begin{equation} \label{theta_eq}
    \theta = \arccos\left(\frac{x(0) - a}{A_0}\right).
\end{equation}

At each iteration, the particle's position is evaluated at a random time. If the personal or global best changes, the corresponding attractors, amplitudes, and phases are recalculated with the clock reset.

Simply limiting energy reduction is not enough to sustain an effective search. When a particle remains in a suboptimal region for too long without updates to the global best, its amplitude may decay to the point where it can no longer explore the space between its personal and global best positions.
This effectively traps the particle, preventing further contributions to the optimization process.
While reducing the damping parameter \(\lambda\) prolong exploration, it may also slow convergence.
To address this, we introduce a threshold amplitude \(A_{\text{th}}\) that ensures particles retain enough energy for continued movement, even in stagnating regions.

The threshold amplitude is defined as proportional to the distance between the personal and global best positions:
\begin{equation} \label{th_A}
    A_{\text{th}} = \frac{|p_{j,d} - g_d|}{2} \cdot m,
\end{equation}
where \(m\) is a tunable scaling factor. The amplitude at each iteration is updated using
\begin{equation}
    (A_0)_{i+1} = \max\left((A_0)_i, A_{\text{re}}, A_{\text{th}}\right),
\end{equation}
where \(A_{\text{re}}\) is the recalculated amplitude following an attractor update. This formulation guarantees that the particle retains sufficient energy to explore meaningfully, enhancing both exploration and convergence robustness.

\begin{figure}[h]
\centering
\includegraphics[width=0.85\linewidth]{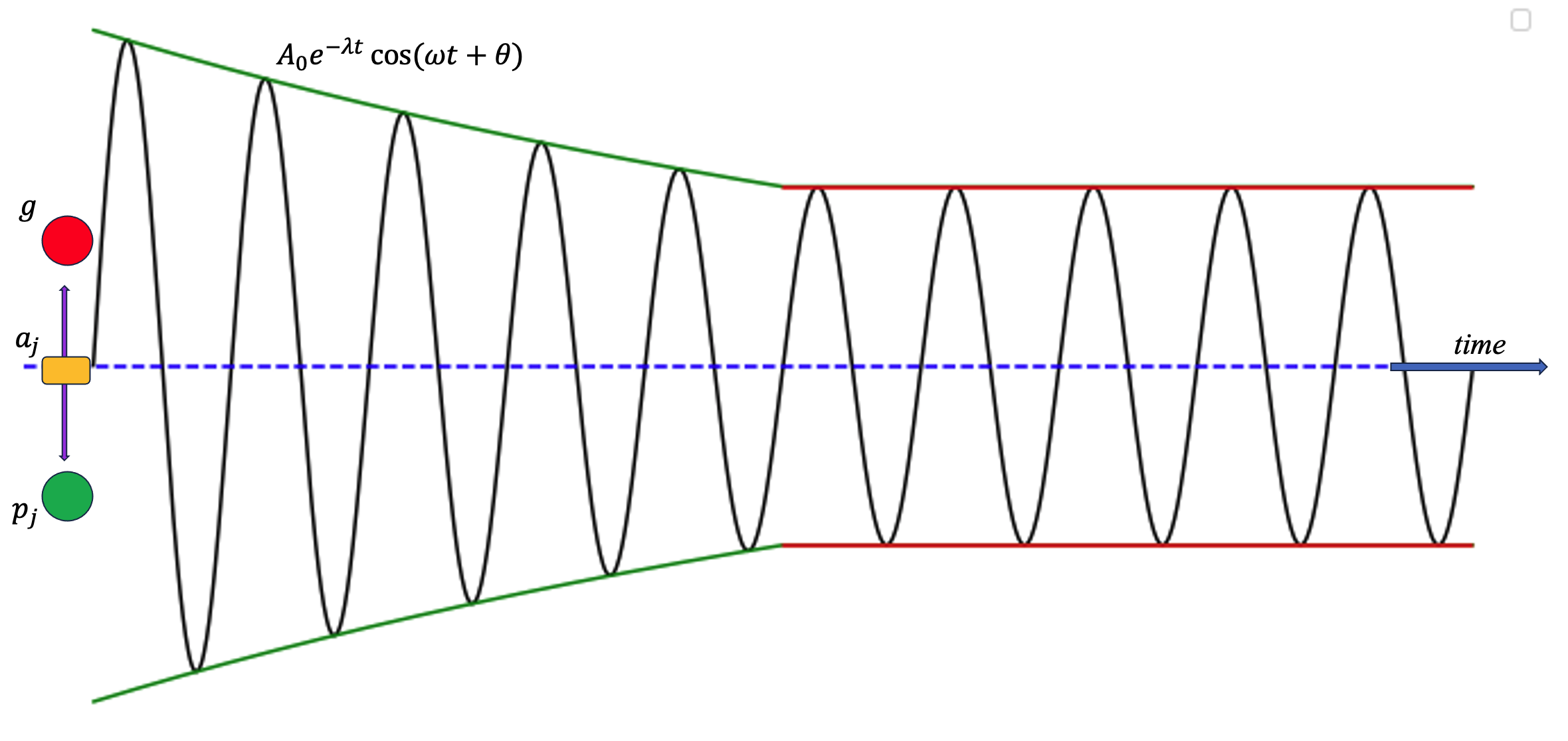}
\caption{One-dimensional illustration of the HOPSO algorithm. A particle oscillates around an attractor $a_j$, which is computed as a weighted average of its personal best position $p_j$ and the global best $g$. The trajectory follows a damped harmonic motion with exponentially decaying amplitude. The amplitude decay is halted when it reaches a threshold proportional to the attractor distance (here, approximately $m = 2.05$ times the attractor distance).}
\label{fig:hopso}
\end{figure}

A visualization of the HOPSO's behavior can be seen in Fig.~\ref{fig:hopso}, which shows how a particle in one dimension evolves over time around its attractor via damped oscillations.

\end{appendices}



\end{document}